\documentclass[times, review, 10pt]{elsarticle}

\usepackage{amssymb}
\usepackage{amsmath}

\usepackage{cleveref}
\usepackage{multirow}
\usepackage{xcolor}
\usepackage{booktabs}
\usepackage{setspace}
\usepackage{url}

\begin{document}

\begin{frontmatter}



\title{P$^3$Net: Progressive and Periodic Perturbation for Semi-Supervised Medical Image Segmentation}

\cortext[cor1]{Corresponding Author
}
\author[Dalian University of Technology]{Zhenyan Yao}
\ead{yzy@mail.dlut.edu.cn}
\author[Dalian University of Technology-2,Key Laboratory for Ubiquitous Network and Service Software of Liaoning Province]{Miao Zhang}
\ead{miaozhang@dlut.edu.cn}
\author[Dalian University of Technology]{Lanhu Wu}
\ead{lanhoong0406@gmail.com}

\author[Dalian University of Technology]{Yongri Piao\corref{cor1}}
\ead{yrpiao@dlut.edu.cn}
\author[Zhongshan]{Feng Tian}
\ead{massurm@163.com}
\author[Zhongshan]{Weibing Sun}
\ead{massurm@163.com}
\author[Dalian University of Technology]{Huchuan Lu}
\ead{lhchuan@dlut.edu.cn}

\affiliation[Dalian University of Technology]{organization={School of Information and Communication Engineering},
            addressline={Dalian University of Technology}, 
            city={Dalian},
            postcode={116024}, 
            state={Liaoning Province},
            country={China}}
\affiliation[Dalian University of Technology-2]{organization={DUT-RU International School of Information Science and Engineering},
            addressline={Dalian University of Technology}, 
            city={Dalian},
            postcode={116024}, 
            state={Liaoning Province},
            country={China}}

\affiliation[Key Laboratory for Ubiquitous Network and Service Software of Liaoning Province]{organization={Key Laboratory for Ubiquitous Network and Service Software of Liaoning Province
},
            country={China}}
\affiliation[Zhongshan]{organization={Affiliated Zhongshan Hospital of Dalian University,},
            addressline={Urology}, 
            city={Dalian},
            country={China}}

\begin{abstract}
Perturbation with diverse unlabeled data has proven beneficial for semi-supervised medical image segmentation (SSMIS). While many works have successfully used various perturbation techniques, a deeper understanding of learning perturbations is needed. Excessive or inappropriate perturbation can have negative effects, so we aim to address two challenges: how to use perturbation mechanisms to guide the learning of unlabeled data through labeled data, and how to ensure accurate predictions in boundary regions.
Inspired by human progressive and periodic learning, we propose a progressive and periodic perturbation mechanism (P$^3$M) and a boundary-focused loss. P$^3$M enables dynamic adjustment of perturbations, allowing the model to gradually learn them. Our boundary-focused loss encourages the model to concentrate on boundary regions, enhancing sensitivity to intricate details and ensuring accurate predictions. Experimental results demonstrate that our method achieves state-of-the-art performance on two 2D and 3D datasets. Moreover, P$^3$M is extendable to other methods, and the proposed loss serves as a universal tool for improving existing methods, highlighting the scalability and applicability of our approach.

\end{abstract}



\end{frontmatter}


\section{Introduction}
\label{sec:intro}

\quad Precise segmentation of medical images offers valuable insights to clinicians, assistsing accurate diagnosis and formulating effective treatment planning  \cite{de2018clinically,ouyang2020video}. Deep learning models have exhibited substantial promise in achieving accurate segmentation results in medical images. This progress is attributed, in part, to the availability of extensive labeled datasets. 
However,the acquisition on labeled data in the medical field poses a significant challenge due to the necessity of expensive equipment for data collection. The emergence of semi-supervised  medical image segmentation (SSMIS)  \cite{bai2023bidirectional,wang2023mcf,basak2023pseudo,cai2023orthogonal,yu2019uncertainty,li2020shape,luo2021semi,chen2023magicnet} has attracted considerable attention in the medical image segmentation for its ability to significantly reduce reliance on labeled data. 

In numerous recent studies, semi-supervised  medical image segmentation can be categorized into two main classes: pseudo-labeling  \cite{chaitanya2023local,seibold2022reference,xie2021intra,yao2022enhancing,tarvainen2017mean} and consistency regularization  \cite{chen2021adaptive,li2020transformation,luo2021semi,shi2021inconsistency}. 
{The shapes and sizes of organs presented in medical images may vary at different angles. Although pseudo-labeling methods have made some improvements in this regard, they may still struggle to adapt well to unseen or diverse medical images.  Consistency regularization is also applied in medical image segmentation. The success lies in that consistency regularization, independent of pseudo-label generation, can encourage the model to learn more generalized and capable feature representations, thus performing better on unseen data.}{One way to perturb input data is a crucial factor affecting SSMIS.}
There are various perturbation techniques  \cite{bai2023bidirectional,kim2022mum,liu2022perturbed,zhao2023instance,yang2023revisiting,zhao2023augmentation,gao2021dsp,olsson2021classmix} enriching unlabeled data and contribute differently to the augmentation process for promoting   robust and generalized learning.

\begin{figure}[t]

  \centering
   \includegraphics[width=1\linewidth]{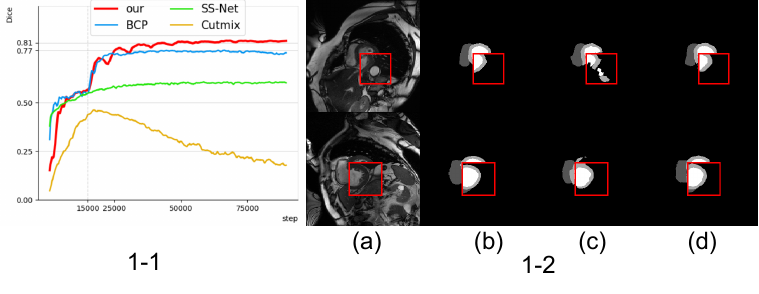}
   \caption{ We assessed the learning progress of unlabeled data every 200 iterations. The Dice coefficient served as our metric to evaluate the amount of information acquired. Fixed interpolation representative: BCP (blue). Random interpolation representative: Cutmix(yellow). Non-interpolation method representative: SSNet (green). }

   \label{fig:figure1}
\end{figure}

It is essential to acknowledge that perturbation techniques can be a double-edged sword while perturbation-methods have demonstrated good performance, excessive or inappropriate perturbations might have adverse effects and overall model performance  \cite{liu2022perturbed}.
Recent perturbation-based methods either design a random-ratio interpolation mechanism  \cite{kim2022mum,gao2021dsp,olsson2021classmix} or fixed-ratio interpolation mechanism  \cite{liu2022perturbed,zhao2023instance,zhao2023augmentation}, aiming to introduce controlled variations.
These controlled variations are aimed at reducing the domain discrepancy between labeled and unlabeled data. However, in this learning process, these methods for reducing domain discrepancy are limited. They do not consider gradually adapting the network to these domain shift transformations.
This may lead to less-effective learning, resulting in suboptimal prediction accuracy. BCP as a representative of the fixed interpolation approach  \cite{bai2023bidirectional} addresses this issue to some extent. However, we can still observe a marginal declining trend in performance as shown in  Fig.\ref{fig:figure1}-1. In addition, the situation becomes even more severe,when using a random interpolation approach (Cutmix  \cite{yun2019cutmix}) represented by yellow line as in Fig. \ref{fig:figure1}-1. Those methods can inevitably lead to over-perturb certain difficult-to-train instances. Such over-perturbations exceed the generalization of the model and impede effective learning \cite{liu2022perturbed}. In addition,  Interpolation-based perturbation methods commonly encounter the challenge of inaccurate boundaries, This stems from the absence of coherent content in the cropped region, especially in areas close to the crop boundaries. Consequently, the predictions may manifest inaccuracies in their boundary as shown in Fig. \ref{fig:figure1}-2.

Based on the above observation, we argue that the perturbation-based methods should be adjusted to form a gradual and intentional learning mechanism for better adapting to SSMIS. 
From the perspective of education and cognitive psychology  \cite{hayes2020remind}, human learning is a progressive and
periodic endeavor.

Progressive learning refers to a gradual process of acquiring new information, and refining understanding over time while periodic learning involves consolidation, reflection, or even temporary setbacks that may occur intermittently. In order to better dynamically adjust perturbation to conform to human learning mechanism, our first challenge toward this goal is how do we devise a perturbation mechanism that mirrors the characteristics of the gradual and cyclical stages of learning. The second challenge is in the design of the loss that enforces the model to focus on the boundary regions, aiming to enhance the model's sensitivity to intricate details and ensure accurate predictions in boundary regions. The key factor contributing to the effectiveness of our approach lies in designing a dynamic perturbation adjustment strategy that mimics the gradual and cyclical learning processes of humans.

Benefiting from progressive and  periodic perturbation and boundary-focused loss design, we obtain state-of-the-art (SOTA) performance on two 3D datasets (LA and Pancreas-NIH) and two 2D datasets (ACDC and Decathlon Prostate). For example, our method obtains a high Dice of 79.69\% with only 6\% labeled data
on Pancreas-NIH, which is 14.25\% higher than the previous SOTA. Our main contributions are summarized as follows:
\begin{itemize} 
\item We propose a progressive and periodic perturbation mechanism ({P$^3$M}). {P$^3$M} systematically regulates interpolated perturbations, acting as a bridge between labeled and unlabeled data to transform positive guidance in a more coherent manner aligned with human learning principles, thereby, enhancing the models' learning ability from unlabeled data.
\item We introduce a boundary-focused loss to improve the model's ability to predict edges by enhancing its confidence.
In doing so, the proposed loss not only assists the model in penalizing errors but also places a heightened emphasis on regions near object boundaries.
\item We assess the scalability of {P$^3$M} through various experiments. We extend the {P$^3$M}  to existing interpolation perturbation  methods, resulting in significant performance improvements using 5\% labeled data  on the ACDC dataset, enhancing  Classmix by 28.30\% in Dice score.
\item Our proposed  boundary-focused loss can be widely used as a universal  tool to improve existing perturbation methods. Compared to the original method, the performance of the method adopting interpolation-based perturbation loss achieves an improvement (5.55\% for MUN) in terms of Dice score using 5\% labeled data on the ACDC dataset.
\end{itemize}

\section{Related Work}
\label{sec:related_work}

\subsection{Semi-supervised Medical Image Segmentation}
In the field of SSMIS, significant progress has been made to enhance performance by addressing practical challenges such as labeling costs and raw data acquisition. This progress is achieved by utilizing a limited amount of labeled data with a wealth of unlabeled data, thereby leveraging the information to compensate for the scarcity of labeled data. Most SSMIS methods fall into two categories: consistency-based methods  \cite{chen2021adaptive,li2020transformation,luo2021semi,shi2021inconsistency} and pseudo-labeling methods  \cite{chaitanya2023local,seibold2022reference,xie2021intra,yao2022enhancing,tarvainen2017mean}.
Consistency-based methods aim to ensure consistent predictions for differently augmented images, enhancing the model's robustness and generalization capability. Pseudo-labeling methods, on the other hand, utilize highly confident labels from a teacher network to train a student network, enabling semi-supervised learning. Among these methods, the mean-teacher framework stands out as a widely adopted architecture, with various extensions explored in the literature.
For example, UA-MT   \cite{yu2019uncertainty} employs uncertainty information to instruct the student network to learn from reliable targets provided by the teacher network, thereby improving model robustness and generalization performance. SASSNet  \cite{li2020shape} introduces an edge distance function, utilizes unlabeled data to impose geometric constraints, and constructs a shape-aware semi-supervised segmentation network with GAN, enhancing the segmentation capability for complex shapes. DTC   \cite{luo2021semi} utilizes a dual-task deep network to jointly predict pixel-level segmentation and a perceptual geometric level set representation, enhancing image understanding and representation capability. Wang  \cite{wang2021tripled} introduces multi-task learning in the MT framework, utilizing triple uncertainty to guide the student model for more reliable predictions. 
Huang  \cite{huang2022semi} suggests a dual-phase approach to neuron segmentation, leveraging consistency in pixel-level predictions between unlabeled samples and their perturbed counterparts. 
DACS  \cite{tranheden2021dacs} proposes a unidirectional interpolation strategy, interpolating from labeled to unlabeled data. In contrast, DSP  \cite{gao2021dsp} introduces a bidirectional interpolation strategy, involving both directions: interpolating from labeled to unlabeled data and vice versa. Their methods enrich and refine SSMIS techniques in different directions.
Overall, significant progress has been achieved in SSMIS by synergizing a limited amount of labeled data with a wealth of unlabeled data, playing a crucial role in the field of image segmentation.
\subsection{Interpolation-based Perturbation Regularization}
IPR is a method to achieve high performance in deep learning networks by preprocessing inputs without injecting noise, and it has only recently received active research attention.
 It augments training data by interpolating original samples with an inductive bias, ensuring the linear combination of two originals closely resembles the output of the interpolated sample. Initially proposed for data augmentation, IPR techniques like Mixup  \cite{zhang2017mixup}, CutMix  \cite{yun2019cutmix}, Mosaic \cite{ge2021yolox}, and Cutout  \cite{devries2017improved} create synthetic training samples. 
However, in recent works, IPR is typically employed in SSLMS learning to better leverage unlabeled data, thereby improving model performance and generalization. ICT  \cite{verma2022interpolation} trains a network using a consistency loss that compares interpolated predictions of two unlabeled samples with that of an interpolated sample. MixMatch  \cite{berthelot2019mixmatch} and ReMixMatch  \cite{berthelot2019remixmatch} generate a pseudo label from multiple views of a single unlabeled image, training it with Mixup  \cite{zhang2017mixup} alongside labeled samples. Additionally, semi-supervised learning in semantic segmentation is extended by   \cite{french2019semi,kim2020structured}. using mixed images from CutMix  \cite{yun2019cutmix}  and the training mechanism of ICT  \cite{verma2022interpolation}. MUM  \cite{kim2022mum} creates mixed input image tiles and reconstructs them in the feature space, while BCP  \cite{bai2023bidirectional} employs bidirectional copy-pasting of labeled and unlabeled data in a simple mean teacher architecture. However, excessive or inappropriate interpolation techniques might introduce unwanted noise and distortions, adversely impacting the learning process and overall model performance. So we propose a progressive and periodic perturbation mechanism.

\section{Method}
\label{sec:method}
\begin{figure*}
  \centering
    \includegraphics[width=1\linewidth]{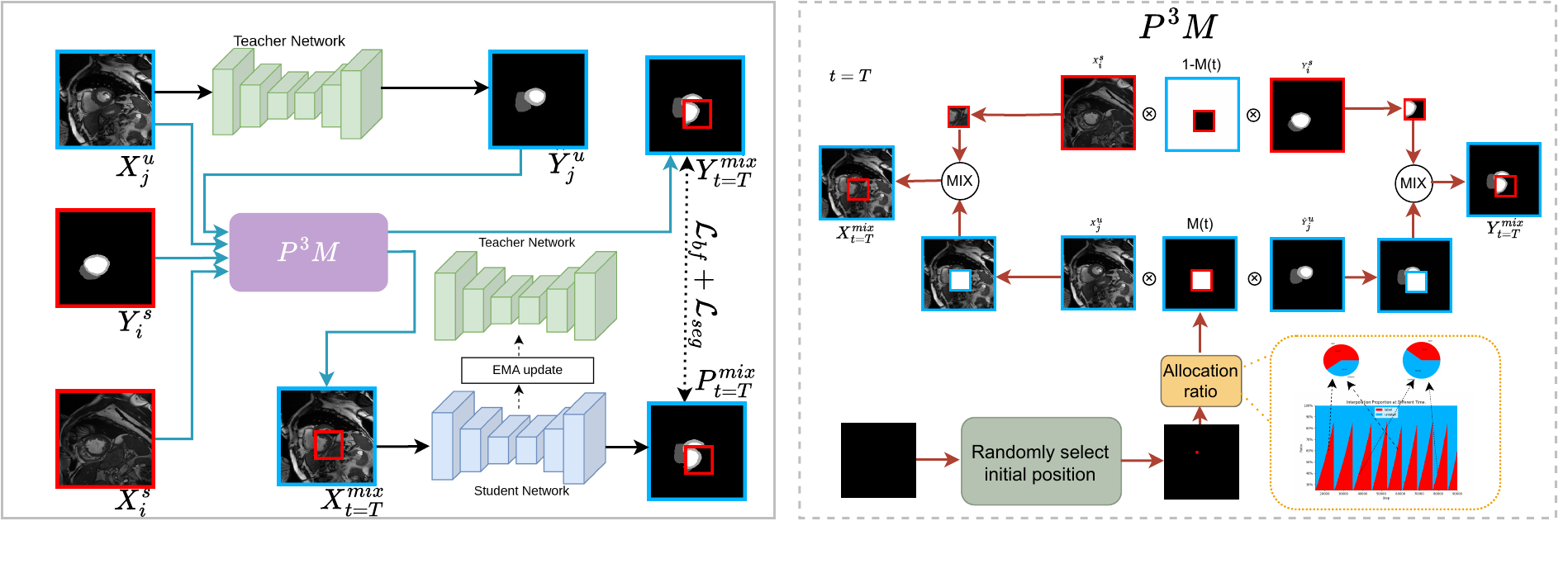}

    \caption{The whole pipeline of our proposed P$^3$Net (left) and progressive  and periodic perturbation mechanism (right) in the second stage.}
    \label{fig:main}
\end{figure*}
\subsection{ Overview}
\label{sec: Overview}
Following the SSMIS  \cite{tarvainen2017mean,yu2019uncertainty}, we aim to train the segmentation encoder and decoder on both labeled and unlabeled data simultaneously. 
Without introducing extra network components,  {P$^3$M} leverages a progressively changing interpolation ratio to enable the network for better utilizing unlabeled information, thereby enhancing performance and generalization capabilities. Our work is applicable to both 2D and 3D datasets. In the following description, we  explain our methods  in 2D inputs.

The objective of SSLMS is to enhance the generalization  of a segmentation model by efficiently harnessing the power of a labeled training set $D_s = {\{(X_i^s, Y_i^s)\}}^{N_s}_{i=1}$ and a substantial unlabeled training set $D_u = {\{X_j^u\}}^{N_u}_{i=1}$. Here, $x$ and $y$ respectively denote an image and its corresponding label.  $N_s$ represents the number of labeled samples where $N_u$ represents the number of unlabeled samples and $N_u$ significantly exceeds $N_s$. 
We adopt an ensemble of teacher models $\mathcal{F}_t(X, {\theta}_T)$, with each model representing a temporal ensemble of the student model $\mathcal{F}_s(X, {\theta}_S)$  \cite{tarvainen2017mean}. This approach is chosen as our baseline. Following the baseline, we first build the teacher network via EMA:
\begin{equation}
\label{eq:ema}
    {\theta}_T^{t+1}= {\theta}_T^t \cdot \delta + {\theta}_S \cdot (1 - \delta),
\end{equation}
where $\theta_T^t, {\theta}_S$ and ${\delta}$ denote the weights of the teacher at step $t$, the weights of the student, and EMA decay rate, respectively. We set  ${\delta=0.99}$.

Our training mechanism is divided into two steps. At the pre-warm step, we train the student network with the pseudo labels generated by the teacher network. 
We utilize $D_s$ along with pseudo-labels $\hat{Y}^u_j$ generated from $D_u$ using a teacher network to train the student network. 
The overall loss in the pre-warm step is formulated as:

\begin{gather}
    \mathcal{L}^{step1}=\mathcal{L}^{step1}_s + \lambda  \mathcal{L}^{step1}_u, \\
    \begin{split}
    \mathcal{L}^{step1}_s &= \mathcal{L}^{step1}_{dice}(\mathcal{F}_s(X_i^s, {\theta}_S),Y_i^s) + 
    \mathcal{L}^{step1}_{ce}(\mathcal{F}_s(X_i^s, {\theta}_S),Y_i^s),\\
    \end{split}\\
     \begin{split}
    \mathcal{L}^{step1}_u &= \mathcal{L}^{step1}_{dice}(\mathcal{F}_s(X_j^u, {\theta}_T),\hat{Y}^u_j) 
    \mathcal{L}^{step1}_{ce}(\mathcal{F}_s(X_j^u, {\theta}_T),\hat{Y}^u_j),
    \end{split}
\end{gather}
where $\mathcal{L}^{step1}_{dice}$ and $\mathcal{L}^{step1}_{ce}$ represents the Dice loss and Cross-entropy loss in the pre-warm stage, 
and $\lambda$ is the balancing weight for the unsupervised loss. We will introduce the second stage in \cref{sec:Progressive and Periodic Perturbation Mechanism}.

\subsection{ Progressive  and Periodic Perturbation Mechanism}
\label{sec:Progressive and Periodic Perturbation Mechanism}
In SSMIS, excessive or inappropriate perturbations may lead to the model overfitting to the noise, thereby reducing performance on unseen data. Therefore, reasonable perturbations are crucial for the effective training and generalization of the model. Human  learning \cite{hayes2020remind} is a gradual and systematic process. It involves transitioning from known knowledge to the learning of unknown concepts through repeated practice. Similarly, our model's learning should also gradually shift from unsupervised to supervised learning, incorporating periodic training to adapt to the distribution differences between labeled and unlabeled data. 
To this end, we design a progressive and periodic perturbation mechanism which enable the perturbations to be a positive guidance. in a manner aligned with human learning principles.

To mimic the progressive leaning process \cite{wozniak1994optimization}, we design a progressively increasing curve function $\alpha_1(iter)$ to gradually guide the network for learning more from unlabeled data:
\begin{equation}
\label{eq:alpha_1}
    \alpha_1(iter_1) =exp(\frac{iter_1}{S})+\gamma,
\end{equation}
where $iter_1$ represents the current iteration step, $\alpha_1(iter)$ represents the proportion of labeled data in the interpolation at iteration $iter$, specifically the area occupied by unlabeled data after using CutMix. $\gamma$ is an adjustment factor. 
In addition, in the context of human learning, $S$ represents the forgetting rate \cite{wozniak1994optimization} \cite{memorymemory}, while in the model training process, we interpret it as the model's ability to reflect on and incorporate new information. $\gamma$ and $S$ are obtained by solving the Eq. \ref{eq:alpha_1} while $\alpha_1(0)=0.25$ and  $\alpha_1(T)=0.9$. Here, we want to emphasize that a reasonable upper and lower limit is far more important than the function itself. Excessive interpolation may introduce unnecessary noise, exceeding the model's generalization capabilities, while insufficient interpolation may hinder the extraction of desired learning information from unlabeled data. The focus of our method is not on this specific curve function, but on the concept of a progressive and periodic perturbation mechanism, which will be analyzed in detail in our ablation experiments.

To mimic the constant review that occurs in human learning due to the phenomenon of forgetting, we introduce the concept of periodic learning. By using periodic perturbations, we help the network adapt to the differences in data distribution between labeled and unlabeled data, thereby mitigating the negative effects of excessive or inappropriate perturbations:
\begin{equation}
\label{eq:zhouqi}
     \alpha(iter) =\alpha_1(iter-n\cdot T),
\end{equation}
and $n$ is the number of current cycles. $T$ is the period of the curve.

In the second stage, as shown in the Fig. \ref{fig:main}, we first input the $X_j^u$ into the teacher network,  their probability maps $P_j^u$ are computed. The initial pseudo-label $\hat{Y}^u_j$ is determined by taking a common threshold 0.5 on $P_j^u$  for binary segmentation tasks, or taking argmax operation for multi-class segmentation tasks.

To conduct interpolation between a pair of images $X_i^s$ and $X_j^u$. We first generate a mask $\mathcal{M}_c(t) \in {\{0,1\}}^{W \times H}$ with all elements set to 1, indicating whether the pixel comes from the foreground (0) or the background (1) image. And $t$ denotes the current moment. We randomly select a pixel and set it at the left upper corner coordinates of the box which is $(w_0,h_0)$. The size of the zero-value region is $ ({\alpha W\times\alpha H})$, so the right lower corner is at  $ {(w_0+\alpha W,h_0+\alpha H)}$. We apply the final mixing step between labeled and unlabeled images to generate the  $X^{mix}_{t=iter}$ at time step $iter$ :
\begin{equation}
\label{eq:M_c_label}
    X^{mix}_{t=iter} = X^u_j  \odot \mathcal{M}_c(t) + X^s_i  \odot (1-\mathcal{M}_c(t)),
\end{equation}
where $\odot$ means element-wise multiplication. Then, the mixed pseudo-labels at time step $iter$ is generated in the same manner:
\begin{equation}
\label{eq:M_c_gt}
    Y^{mix}_{t=iter} =\hat{Y}^u_j  \odot \mathcal{M}_c(t) + Y^s_i  \odot (1-\mathcal{M}_c(t)),
\end{equation}
$Y^{mix}_{t=iter} $ is used to supervise the Student network.

\subsection{Boundary-Focused Loss}
\label{sec:Loss of Interpolation-based Regularization}
After obtaining the interpolated images, we may undergo a loss of interpolated boundary texture and shape information. To address this issue, we propose a boundary-focused loss. Specifically, We propose a weighted  cross entropy (wCE) loss and weighted dice loss (wDICE) :
\begin{equation}
\label{eq:wce}
    \mathcal{L}_{wce} = \frac{\sum\limits_{i=1}^{H} \sum\limits_{j=1}^{W}(1+\mu_{i,j})\sum\limits_{l=0}^{n-1}\textbf{1}(y_{ij}^s=l)log(\textbf{Pr}(p_{ij}^s=l|\Psi))}{\sum\limits_{i=1}^{H} \sum\limits_{j=1}^{W} 1} ,
\end{equation}
\vspace{-0.30cm}
\begin{equation}
\label{eq:wdice}
    \mathcal{L}_{wdice} = 1-  \frac{1}{n}\cdot\frac{2\cdot \sum\limits_{l=0}^{n-1}\sum\limits_{i=1}^{H} \sum\limits_{j=1}^{W}(1+\mu_{i,j})*(y_{ij}^s*p_{i,j}^s)}{\sum\limits_{i=1}^{H} \sum\limits_{j=1}^{W} (y_{ij}^s+p_{i,j}^s)} ,
\end{equation}
where $\textbf{1}(\cdot)$ is the indicator function and The notation $l \in \{0, 1,\cdot\cdot\cdot,n-1\}$ indicates n kinds of labels. 
$p_{i,j}^s$ and $y^s_{j,j}$ are prediction and $Y^{mix}_{t=iter}$ of the pixel at location $(i, j)$ in an image. $\textbf{Pr}(p_{ij}^s=l|\Psi)$ denotes the predicted probability. 
In $ \mathcal{L}_{wce}$ and $ \mathcal{L}_{wdice}$  each pixel is assigned with a weight $\mu_{i,j}$. Hard pixel corresponds to larger  $\mu_{i,j}$ and simple pixel is assigned a smaller one. $\mu_{i,j}$ could be regarded as the indicator of pixel importance, which is calculated according to the difference between the center pixel and its surroundings. 
\vspace{-0.2cm}
\begin{equation}
\label{eq:mu}
    \mu_{i,j} = 1-\frac{{\sum\limits_{ A_{i,j}}}\textbf{1}(y_{i,j}^s=p_{i,j}^s)}{\sum\limits_{A_{i,j}}1},
\end{equation}

where $A_{i,j}$ represents the area that surrounds the pixel $(i, j)$ (pixel $\in \mathcal{M}_l$).  $A_{i,j}$ is the set of pixels in the $5 \times 5$ neighborhood around pixel $(i, j)$. Because interpolation boundaries lack contextual information, this often leads to prediction errors surrounding the interpolation boundary. At this point, the value of $\mu_{i,j}$ tends to be large, thereby intensifying the penalty on the network, thus achieving the purpose of correction.
For all pixels, $\mu_{i,j} \in[0, 1]$. If $\mu_{i,j}$ is large, the pixel at $(i, j)$ is very different from its surroundings. So it is an error-prone pixel  and deserves more attention, indicating that it is concentrated around the edges.

 We generate a  mask $\mathcal{M}_l$ to set all 0. When starting with the upper-left and lower-right coordinates of $\mathcal{M}_c(t)$ as $(h_0, w_0)$ and $(h_1, w_1)$, respectively, we expand outward and shrink inward by $\epsilon$ pixels. The expanded region extends from $(h_0-\epsilon, w_0-\epsilon)$ to $(h_1+\epsilon, w_1+\epsilon)$, while the shrunken region ranges from $(h_0+\epsilon, w_0+\epsilon)$ to $(h_1-\epsilon, w_1-\epsilon)$. The final region where we set to 1 is the expanded region minus the shrunken region. 
So our  loss of  interpolation-based perturbation loss:
\begin{align}
\label{eq:bf}
\mathcal{L}_{bf} = \mathcal{L}_{wce}(\mathcal{F}_s( X^{mix}_{t=iter}, {\theta}_S), Y^{mix}_{t=iter}) 
+  \mathcal{L}_{wdice}(\mathcal{F}_s( X^{mix}_{t=iter}, {\theta}_S), Y^{mix}_{t=iter}),
\end{align}
and loss in other regions :
\begin{align}
\label{eq:seq}
\mathcal{L}_{seg} = \mathcal{L}_{ce}(\mathcal{F}_s( X^{mix}_{t=iter}, {\theta}_S), Y^{mix}_{t=iter}) 
+  \mathcal{L}_{dice}(\mathcal{F}_s( X^{mix}_{t=iter}, {\theta}_S), Y^{mix}_{t=iter}),
\end{align}
Our loss function in the second stage is as follows:
\begin{align}
\label{eq:step2loss}
    \mathcal{L}^{step2}= \mathcal{M}_l \odot \mathcal{L}_{bf}  + (1-\mathcal{M}_l) \odot \mathcal{L}_{seg}.
\end{align}
Afterwards, the teacher network updates parameters using  Eq.\ref{eq:ema}.

\section{Experiments}
\label{sec:experiment}
\subsection{Dataset and Evaluation Metrics}
\begin{table*}[]
\centering

\resizebox{\textwidth}{!}{
\begin{tabular}{cllllllllllllllll}
\hline
\multicolumn{17}{c}{2D dataset}                                                                                                                                                                                                                                                                                                                                                                                                                                         
\\ \hline
\multicolumn{1}{c|}{}  & \multicolumn{8}{c|}{ACDC}  & \multicolumn{8}{c}{Decathlon Prostate}  
\\ \cline{2-17}
\multicolumn{1}{c|}{}  & \multicolumn{4}{c|}{5\%(3 case)} & \multicolumn{4}{c|}{10\%(7 case)}  &\multicolumn{4}{c|}{5\%(23)}   & \multicolumn{4}{c}{10\%(46)} 
\\ \cline{2-17} 
\multicolumn{1}{c|}{\multirow{-3}{*}{Method}} & Dice                         & Jaccard                      & HD95                         & \multicolumn{1}{l|}{ASD}                         & Dice                         & Jaccard                      & HD95                        & \multicolumn{1}{l|}{ASD}                         & Dice                         & Jaccard                      & HD95                        & \multicolumn{1}{l|}{ASD}                         & Dice                         & Jaccard                      & HD95                        & ASD                         \\ \hline
\multicolumn{1}{c|}{UAMT$_{(MICCAI2019)}$}                     & 61.74                        & 49.42                        & 31.49                        & \multicolumn{1}{l|}{10.13}                       & 83.04                        & 72.36                        & 10.14                       & \multicolumn{1}{l|}{2.90}                        & 50.41                        & 18.93                        & 9.46                        & \multicolumn{1}{l|}{3.04}                        & 65.62                        & 34.69                        & 7.04                        & 2.07                        \\
\multicolumn{1}{c|}{SASSNET$_{(MICCAI2020)}$}                  & 63.42                        & 51.39                        & 23.64                        & \multicolumn{1}{l|}{8.46}                        & 84.13                        & 73.89                        & 5.03                        & \multicolumn{1}{l|}{1.41}                        & 52.39                        & 22.11                        & 9.05                        & \multicolumn{1}{l|}{2.78}                        & 66.20                        & 36.07                        & 5.72                        & 1.93                        \\
\multicolumn{1}{c|}{DTC$_{(AAAI2021)}$}                      & 64.25                        & 53.14                        & 8.54                         & \multicolumn{1}{l|}{6.20}                        & 84.59                        & 74.15                        & 8.78                        & \multicolumn{1}{l|}{2.03}                        & 53.10                        & 46.85                        & 8.62                        & \multicolumn{1}{l|}{2.84}                        & 67.86                        & 34.72                        & 5.89                        & 1.78                        \\
\multicolumn{1}{c|}{SSNet$_{(MICCAI2022)}$}                    & 65.82                        & 55.38                        & 6.67                         & \multicolumn{1}{l|}{2.31}                        & 86.78                        & 77.67                        & 6.07                        & \multicolumn{1}{l|}{1.90}                        & 55.08                        & 22.67                        & 8.64                        & \multicolumn{1}{l|}{2.48}                        & 68.70                        & 36.71                        & 6.82                        & 1.97                        \\
\multicolumn{1}{c|}{SLCNet$_{(MICCAI2022)}$}                   & 70.38                        & 57.47                        & 10.13                        & \multicolumn{1}{l|}{2.08}                        & 82.20                        & 71.16                        & 10.10                       & \multicolumn{1}{l|}{1.80}                        & 50.49                        & 22.22                        & 11.20                       & \multicolumn{1}{l|}{4.03}                        & 59.31                        & 31.27                        & 9.63                        & 3.05                        \\
\multicolumn{1}{c|}{SCPNet$_{(MICCAI2023)}$}                   & 71.07                        & 59.73                        & 15.34                        & \multicolumn{1}{l|}{5.04}                        & 87.55                        & 78.67                        & 8.10                        & \multicolumn{1}{l|}{1.96}                        & 59.40                        & 26.49                        & 8.98                        & \multicolumn{1}{l|}{2.84}                        & 69.58                        & 36.33                        & 6.91                        & 2.15                        \\
\multicolumn{1}{c|}{DCNet$_{(MICCAI2023)}$}                    & 71.02                        & 60.10                        & 9.79                         & \multicolumn{1}{l|}{2.45}                        & 89.15                        & 81.62                        & 1.64                        & \multicolumn{1}{l|}{0.58}                        & 58.20                        & 25.90                        & 9.06                        & \multicolumn{1}{l|}{2.92}                        & 72.48                        & 37.97                        & 4.67                        & 1.29                        \\
\multicolumn{1}{c|}{MCF$_{(CVPR2023)}$}                      & 43.36                        & 31.28                        & 68.07                        & \multicolumn{1}{l|}{29.68}                       & 72.39                        & 61.24                        & 9.56                        & \multicolumn{1}{l|}{3.78}                        & 50.49                        & 45.74                        & 7.85                        & \multicolumn{1}{l|}{3.56}                        & 53.84                        & 49.29                        & 5.82                        & 2.57                        \\
\multicolumn{1}{c|}{BCP$_{(CVPR2023)}$}                      & 87.59                        & 78.67                        & {1.91}  & \multicolumn{1}{l|}{0.67}                        & 88.84                        & 80.62                        & 3.98                        & \multicolumn{1}{l|}{1.17}                        & 64.49                        & 37.09                        & 10.03                       & \multicolumn{1}{l|}{5.10}                        & 71.36                        & 38.90                        & 4.94                        & 1.57                        \\
\multicolumn{1}{c|}{ABD$_{(CVPR2024)}$} & 88.96 & 80.70 & 1.57 & \multicolumn{1}{l|}{0.52} & 89.81 & 81.95 & 1.46 & \multicolumn{1}{l|}{0.49} & 64.89 & 45.94 & 7.99 & \multicolumn{1}{l|}{3.84} & 72.54 & 59.34 & 4.58 & 1.63\\

\multicolumn{1}{c|}{GA-BCP$_{(ECCV2024)}$} & 88.24 & 79.60 & 3.91 & \multicolumn{1}{l|}{1.11} & 89.31 & 81.27 & 3.32 & \multicolumn{1}{l|}{1.01} & 64.32 & 45.84 & 7.54 & \multicolumn{1}{l|}{4.03} & 71.15 & 58.64 & 6.04 & 2.48 \\

\multicolumn{1}{c|}{AD-MT$_{(ECCV2024)}$} & 88.75 & 80.41 & \color[HTML]{FF0000}{1.48} & \multicolumn{1}{l|}{\color[HTML]{FF0000}{0.50}} & 89.46 & 81.47 & 1.51 & \multicolumn{1}{l|}{0.44} & 65.02 & 57.84 & 7.69 & \multicolumn{1}{l|}{2.98} & 72.84 & 64.25 & 4.87 & 1.90 \\

\multicolumn{1}{c|}{Ours}                         & {\color[HTML]{FF0000} 89.71} & {\color[HTML]{FF0000} 81.88} & {\color[HTML]{333333} 1.85}  & \multicolumn{1}{l|}{{0.64}} & {\color[HTML]{FF0000} 91.16} & {\color[HTML]{FF0000} 84.15} & {\color[HTML]{FF0000} 1.28} & \multicolumn{1}{l|}{{\color[HTML]{FF0000} 0.36}} & {\color[HTML]{FF0000} 65.63} & {\color[HTML]{FF0000} 58.46} & {\color[HTML]{FF0000} 7.45} & \multicolumn{1}{l|}{{\color[HTML]{FF0000} 2.45}} & {\color[HTML]{FF0000} 74.27} & {\color[HTML]{FF0000} 67.57} & {\color[HTML]{FF0000} 4.49} & {\color[HTML]{FF0000} 1.57} \\
\hline
\multicolumn{1}{c|}{}                         & \multicolumn{8}{c|}{100\%(140case)}                                                                                                                                                                                                                                                          & \multicolumn{8}{c}{100\%(458)}                                                                                                                                                                                                                                         \\ \cline{2-17} 
\multicolumn{1}{c|}{\multirow{-2}{*}{Full}}   & 92.27                        & 85.95                        & 1.08                         & \multicolumn{1}{l|}{0.28}                        & 92.27                        & 85.95                        & 1.08                        & \multicolumn{1}{l|}{0.28}                        & 74.81                        & 68.70                        & 4.41                        & \multicolumn{1}{l|}{1.47}                        & 74.81                        & 68.70                        & 4.41                        & 1.47                        \\ \hline
\multicolumn{1}{l}{}                          &                              &                              &                              &                                                  &                              &                              &                             &                                                  &                              &                              &                             &                                                  &                              &                              &                             &                             \\ \hline
\multicolumn{17}{c}{3D      dataset}                                                                                                                                                                                                                                                                                                                                                                                                                                                                                                                                                                                  \\ \hline
\multicolumn{1}{c|}{}                         & \multicolumn{8}{c|}{LA}                                                                                                                                                                                                                                                                      & \multicolumn{8}{c}{Pancreas-NIH}                                                                                                                                                                                                                                       \\ \cline{2-17} 
\multicolumn{1}{c|}{}                         & \multicolumn{4}{c|}{5\%(4 case)}                                                                                                              & \multicolumn{4}{c|}{10\%(8 case)}                                                                                                            & \multicolumn{4}{c|}{6\%(4 case)}                                                                                                             & \multicolumn{4}{c}{10\%(6 case)}                                                                                        \\ \cline{2-17} 
\multicolumn{1}{c|}{\multirow{-3}{*}{Method}} & Dice                         & Jaccard                      & HD95                         & \multicolumn{1}{l|}{ASD}                         & Dice                         & Jaccard                      & HD95                        & \multicolumn{1}{l|}{ASD}                         & Dice                         & Jaccard                      & HD95                        & \multicolumn{1}{l|}{ASD}                         & Dice                         & Jaccard                      & HD95                        & ASD                         \\ \hline
\multicolumn{1}{c|}{UAMT$_{(MICCAI2019)}$}                     & 78.84                        & 65.65                        & 20.20                        & \multicolumn{1}{l|}{6.48}                        & 82.52                        & 70.95                        & 17.03                       & \multicolumn{1}{l|}{4.94}                        & 42.68                        & 29.28                        & 31.80                       & \multicolumn{1}{l|}{10.94}                       & 51.98                        & 37.44                        & 23.12                       & 8.28                        \\
\multicolumn{1}{c|}{SASSNET$_{(MICCAI2020)}$}                  & 80.68                        & 68.40                        & 26.76                        & \multicolumn{1}{l|}{7.06}                        & 86.80                        & 76.91                        & 14.56                       & \multicolumn{1}{l|}{4.11}                        & 56.06                        & 40.66                        & 26.83                       & \multicolumn{1}{l|}{9.19}                        & 61.42                        & 45.80                        & 21.66                       & 6.56                        \\
\multicolumn{1}{c|}{DTC$_{(DCT2021)}$}                      & 82.28                        & 70.57                        & 16.43                        & \multicolumn{1}{l|}{4.50}                        & 87.42                        & 78.06                        & 8.38                        & \multicolumn{1}{l|}{2.40}                        & 46.41                        & 33.55                        & 45.79                       & \multicolumn{1}{l|}{46.41}                       & 60.06                        & 45.54                        & 34.36                       & 8.59                        \\
\multicolumn{1}{c|}{SSNet$_{(MICCAI2022)}$}                    & 86.33                        & 76.15                        & 9.95                         & \multicolumn{1}{l|}{2.33}                        & 88.54                        & 79.60                        & 7.59                        & \multicolumn{1}{l|}{1.90}                        & 54.19                        & 39.43                        & 20.70                       & \multicolumn{1}{l|}{7.20}                        & 63.38                        & 47.77                        & 21.59                       & 4.65                        \\
\multicolumn{1}{c|}{CAML$_{(MICCAI2023)}$}                     & 87.54                        & 77.95                        & 10.76                        & \multicolumn{1}{l|}{2.58}                        & 89.44                        & 81.01                        & 10.10                       & \multicolumn{1}{l|}{2.59}                        & 56.65                        & 41.65                        & 28.96                       & \multicolumn{1}{l|}{9.74}                        & 70.59                        & 56.00                        & 12.63                       & 3.20                        \\
\multicolumn{1}{c|}{UpCoL$_{(MICCAI2023)}$}                    & 80.85                        & 69.26                        & 17.51                        & \multicolumn{1}{l|}{5.43}                        & 86.89                        & 77.35                        & 14.57                       & \multicolumn{1}{l|}{4.10}                        & 53.14                        & 38.64                        & 39.74                       & \multicolumn{1}{l|}{14.92}                       & 59.92                        & 45.68                        & 26.47                       & 8.93                        \\
\multicolumn{1}{c|}{MagicNet$_{(CVPR2023)}$}                 & 84.22                        & 71.85                        & 12.44                        & \multicolumn{1}{l|}{3.71}                        & 86.83                        & 77.16                        & 14.57                       & \multicolumn{1}{l|}{4.06}                        & 70.06                        & 55.00                        & 16.38                       & \multicolumn{1}{l|}{5.41}                        & 75.32                        & 61.94                        & 12.30                       & 4.54                        \\
\multicolumn{1}{c|}{MCF$_{(CVPR2023)}$}                      & 82.41                        & 70.77                        & 14.61                        & \multicolumn{1}{l|}{4.32}                        & 87.35                        & 78.21                        & 8.38                        & \multicolumn{1}{l|}{2.40}                       & 40.15                        & 27.45                        & 30.38                       & \multicolumn{1}{l|}{12.36}                       & 48.77                        & 33.95                        & 36.83                       & 13.38                       \\
\multicolumn{1}{c|}{BCP$_{(CVPR2023)}$}                      & 88.02                        & 78.72                        & {\color[HTML]{333333} 7.90}  & \multicolumn{1}{l|}{{2.26}} & 89.62                        & 81.31                        & {6.82} & \multicolumn{1}{l|}{{1.81}} & 67.38                        & 52.50                        & 22.85                       & \multicolumn{1}{l|}{5.29}                        & 75.98                        & 62.45                        & 11.77                       & 3.06                        \\

\multicolumn{1}{c|}{CVML$_{(AAAI2024)}$} & \multicolumn{1}{l}{87.63} & 79.04 & \multicolumn{1}{l}{8.92} & \multicolumn{1}{l|}{2.23} & \multicolumn{1}{l}{90.36} & 82.46 & \multicolumn{1}{l}{6.06} & \multicolumn{1}{l|}{1.68} & 66.72 & 50.69 & 22.87 & \multicolumn{1}{l|}{6.24} & 73.54 & 54.28 & 13.67 & 5.64 \\
\multicolumn{1}{c|}{LeFeD$_{(TMI2024)}$} & 88.56                     & 78.10 & 10.62                    & \multicolumn{1}{l|}{3.41 }                    & 90.12                     & 81.78 & 7.02                     & \multicolumn{1}{l|}{1.74 }                    & 63.95 & 49.35 & 20.87 & \multicolumn{1}{l|}{5.64} & 75.35 & 61.38 & 11.47 & 3.42 \\
\multicolumn{1}{c|}{PMT$_{(ECCV2024)}$}  & 89.47                     & 81.04 & \color[HTML]{FF0000}{6.45}                     & \multicolumn{1}{l|}{{\color[HTML]{FF0000} 1.86}}                     & 90.81                     & 83.23 & \color[HTML]{FF0000} 5.61                     & \multicolumn{1}{l|}{\color[HTML]{FF0000} 1.50}                     & 68.32 & 53.04 & 18.06 & \multicolumn{1}{l|}{4.57} & 81.00 & 68.33 & \color[HTML]{FF0000} 6.36  & \color[HTML]{FF0000} 1.62\\

\multicolumn{1}{c|}{Ours}                         & {\color[HTML]{FF0000} 89.57} & {\color[HTML]{FF0000} 81.17} & {7.83}  & \multicolumn{1}{l|}{{\color[HTML]{333333} 2.33}} & {\color[HTML]{FF0000} 90.83} & {\color[HTML]{FF0000} 83.30} & {\color[HTML]{333333} 7.02} & \multicolumn{1}{l|}{{\color[HTML]{333333} 1.92}} & {\color[HTML]{FF0000} 79.69} & {\color[HTML]{FF0000} 66.53} & {\color[HTML]{FF0000} 8.61} & \multicolumn{1}{l|}{{\color[HTML]{FF0000} 2.72}} & {\color[HTML]{FF0000} 81.70} & {\color[HTML]{FF0000} 69.39} & {7.71} & {2.10} \\
\hline
\multicolumn{1}{c|}{}                         & \multicolumn{8}{c|}{100\%(140case)}                                                                                                                                                                                                                                                          & \multicolumn{8}{c}{100\%(46)}                                                                                                               \\ \cline{2-17} 
\multicolumn{1}{c|}{\multirow{-2}{*}{Full}}   & 92.26                        & 85.68                        & 6.21                         & 1.35                                             & 92.26                        & 85.68                        & 6.21                        & \multicolumn{1}{l|}{1.35}                        & 82.31                        & 70.27                        & 7.29                        & 1.76                                             & 82.31                        & 70.27                        & 7.29                        & 1.76                        \\ 
\hline
\end{tabular}
}
\small

\caption{Comparison of our method with state-of-the-art semi-supervised segmentation methods on four datasets. The best results are shown in \textcolor{red}{red}. Higher values for Dice and Jaccard indicate better performance ($\uparrow$), while lower values for HD95 and ASD are desirable ($\downarrow$).}
\label{table:compare with sota}
\end{table*}
\noindent \textbf{Left Atrial Dataset (LA).} The LA dataset \cite{xiong2021global} includes 100 3D gadolinium-enhanced MR imaging volumes
with an isotropic resolution of $0.625 \times 0.625 \times 0.625mm^3$ and the corresponding ground truth labels. We follow the works \cite{yu2019uncertainty,wu2022exploring} to split the dataset into 80-20 as the training and test sets, respectively.
 During training, the training volumes are randomly cropped to $112\times112\times80$ as the model input. During inference, a sliding window of the same size is used to obtain segmentation results with a stride of $18\times 18 \times4$.\\
 \textbf{NIH pancreas Dataset.} A publicly available NIH Pancreas dataset \cite{roth2015deeporgan} provides 82 contrast-enhanced abdominal 3D CT volumes with manual annotation.  The size of each CT volume is $512 \times 512 \times D$, where $D\in [181,466]$. The data split \cite{shi2021inconsistency,wang2023mcf} is fixed with 62-20 as the training, validation, and test sets, respectively. During training, the training volumes are randomly cropped to $96\times 96\times 96$ as the model input. During inference, a sliding window of the same size is used to obtain segmentation results with a stride of $16\times 16 \times4$ training, validation, and testing.\\
\textbf{ACDC Dataset.} The ACDC \cite{bernard2018deep} is a cardiac MRI dataset  that contains 100 short axis cine-MRIs, captured using 3T and 1.5T machines, and contains expert annotations for three classes: left and right ventricle (LV, RV), and myocardium (MYO). 
We follow the works \cite{wu2022cross,ssl4mis2020} to split the dataset into 70-10-20 as the training, validation, and test sets, respectively.\\
\textbf{Medical Segmentation Decathlon Prostate Dataset.} This dataset \cite{antonelli2022medical} includes 48 multi-parametric MRI studies provided by Radboud University, with 32 MRIs labeled. The ground-truth annotations cover two prostate structures: the peripheral zone (PZ) and central gland (CG). We followed the works \cite{wang2023cat} to split the dataset into 24-8 as the training and test sets, respectively. \\
 \textbf{Metrics.} We use four metrics to evaluate model performance, including regional sensitive metrics: Dice similarity coefficient (Dice), Jaccard similarity coefficient (Jaccard), and edge sensitive metrics: 95\% Hausdorff Distance (95HD) and Average Surface Distance (ASD).\\

 \subsection{Implementation Details.} 
 We set $T=8000$, $\epsilon=13$. We follow  \cite{laine2016temporal} to control the weight $\lambda(t)=0.1*e^{-5(1-\frac{iter}{max_iter})}$. We conduct all experiments on an NVIDIA 4090 GPU with fixed random seeds. We use Adam with  the initial learning rate $3e-4$ and weight decay of 0.0001 as optimizer.  Following previous papers  \cite{chen2021semi,he2021re,ouali2020semi}, we utilize the following polynomial learning-rate decay: ${(1-\frac{iter}{max_iter})}^{0.9}$.
 Semi-supervised experiments of different labeled ratios (\textit{i.e.}, 5\% and 10\%) are carried out. For the 3D dataset, the backbone is configured as a 3D V-Net  \cite{milletari2016fully}. And the batch size, the pre-warm step iterations and the second step iterations are set as 4, 5k and 55k, respectively. For the 2D dataset, the backbone is configured as a U-Net  \cite{ronneberger2015u}. The batch size, the pre-warm step iterations and the second step iterations are set as 16, 15k and 75k, respectively. 
 Furthermore,  following CoraNet  \cite{shi2021inconsistency}, we set the learning rate to 0.001 and semi-supervised experiments of different labeled ratios (\textit{i.e.}, 6\% and 10\%) are carried out on the NIH pancreas dataset.

\subsection{ Comparison with State-of-the-Art Methods}
In comparing our method with state-of-the-art methods, we thoroughly evaluated its performance on both 2D and 3D datasets. This comprehensive analysis gauges the versatility and effectiveness of our approach across various medical imaging scenarios.

\begin{figure*}
  \centering
     \includegraphics[width=1\linewidth]{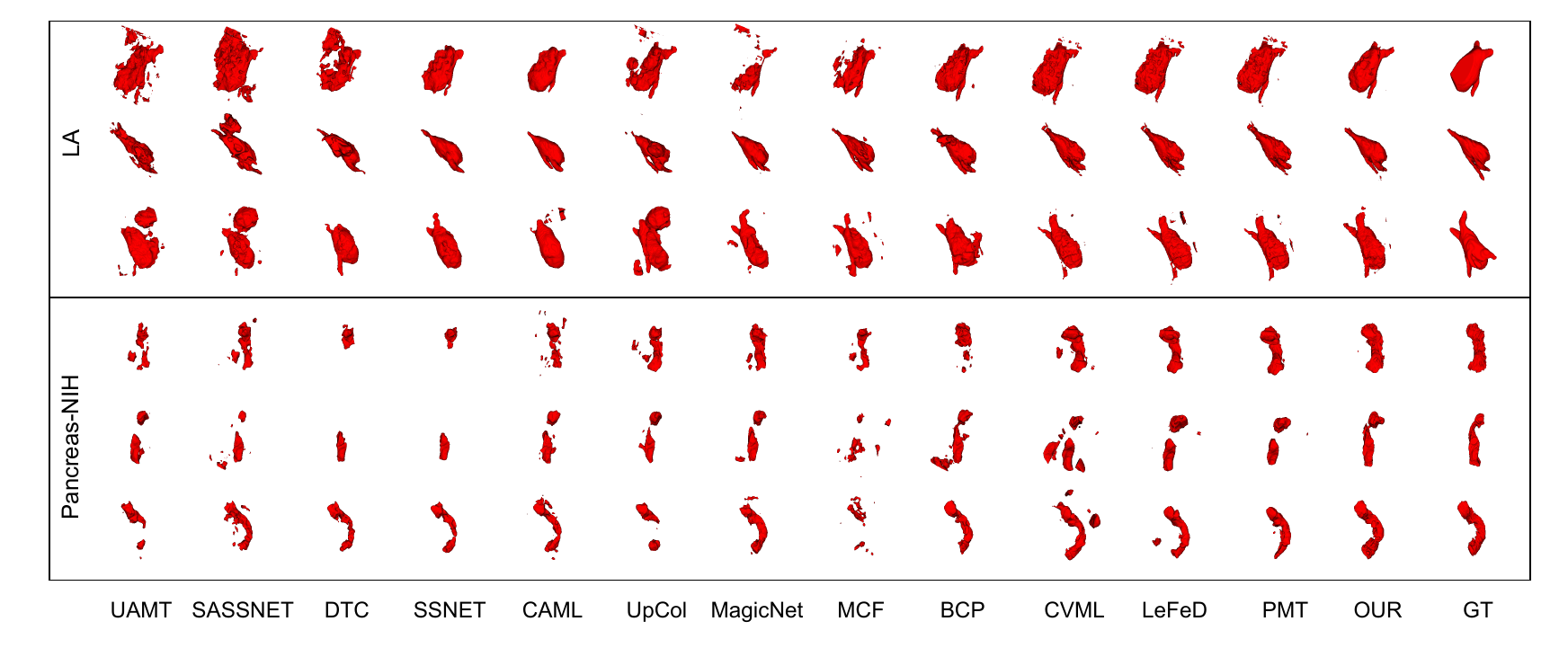}
    \caption{Visualization of 3D datasets LA with 5\% labeled data and Pancreas-NIH with 6\% labeled data and ground truth.(best viewed by zoom-in on screen).}
    \label{fig:3D}
\end{figure*}

\textbf{3D Dataset: }We compare our framework on the 3D datasets (LA and Pancreas-NIH) with various methods: UAMT  \cite{yu2019uncertainty}, SASSNet  \cite{li2020shape}, DTC  \cite{luo2021semi}, SS-Net  \cite{wu2022exploring}, CAML  \cite{gao2023correlation}, UpCoL  \cite{lu2023upcol}, MagicNet  \cite{chen2023magicnet}, MCF  \cite{wang2023mcf}, BCP  \cite{bai2023bidirectional}, CVML \cite{huang2024combinatorial}, LeFeD \cite{10619990}, PMT \cite{gao2025pmt}. As shown in Tab. \ref{table:compare with sota}, we conduct semi-supervised experiments with different labeled ratios on each dataset. As we can see, our method achieves the best performance on all four evaluation metrics, outperforming other methods by a big margin in Pancreas-NIH. In the Pancreas-NIH dataset, our model's performance improves by 11.37\% in Dice when utilizing only 6\% of labeled data. Furthermore, with just 6\% labeled data, our model still outperforms most models that use 10\% labeled data.
Furthermore, we also achieve significant improvements in Dice and Jaccard scores on the LA dataset.  Our methods can also slightly outperform or remain largely consistent with theirs without any explicit boundary or shape constraints during training on HD95 and ASD.  Fig. \ref{fig:3D}  shows visual comparisons to demonstrate the superiority of our proposed approach in an intuitive way. As we can see, the target organs can be conducted more effectively. Our method also can accurately predict  the pulmonary veins on the LA dataset and achieve the structure of the pancreas completely on the NIH dataset where other methods failed to successfully predict them. \\
\begin{figure*}
  \centering
    \includegraphics[width=1\linewidth]{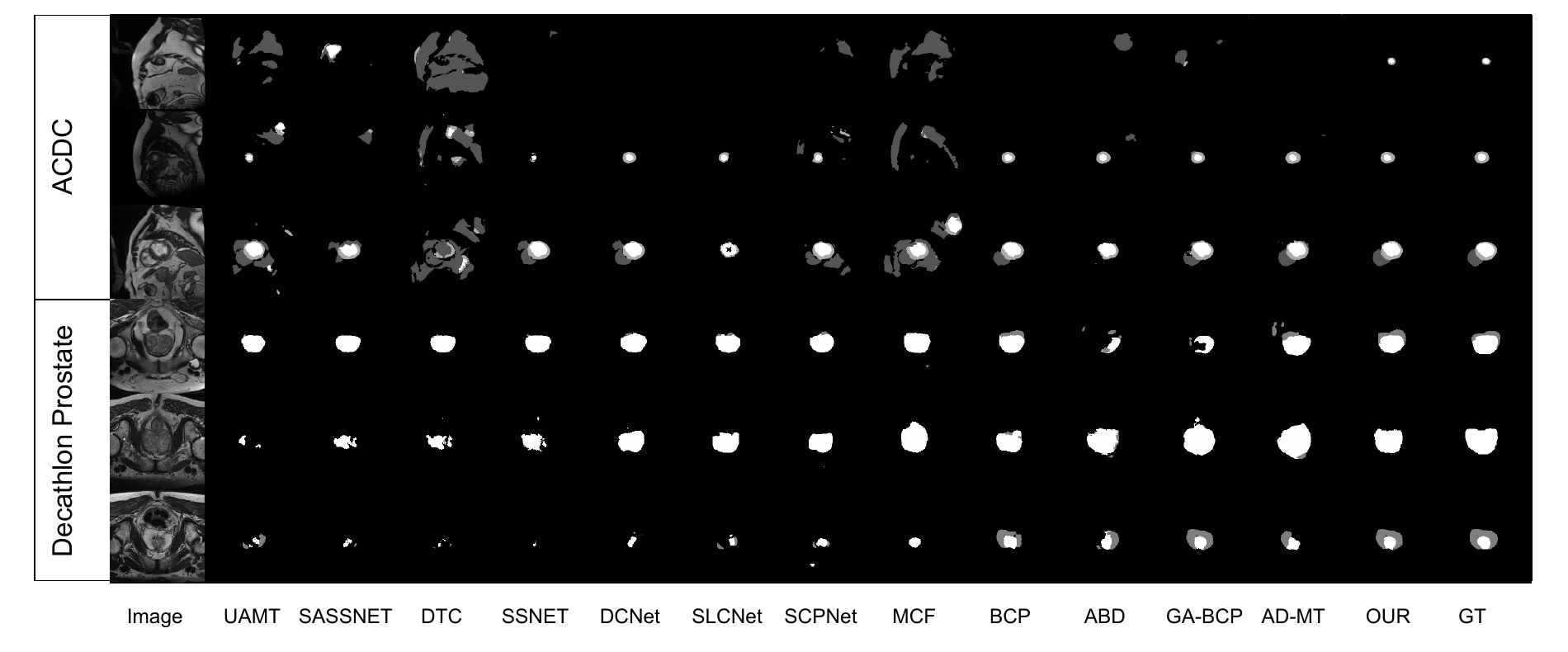}
    \caption{Visualization of 3D datasets LA with 5\% labeled data and Pancreas-NIH with 6\% labeled data and ground truth.(best viewed by zoom-in on screen).}
    \label{fig:2D}
\end{figure*}
\textbf{2D Dataset: }We also conduct comparisons with various methods on the two 2D datasets (ACDC and Decathlon Prostate): UAMT  \cite{yu2019uncertainty}, SASSNet  \cite{li2020shape}, DTC  \cite{luo2021semi}, SS-Net  \cite{wu2022exploring}, DCNet  \cite{chen2023decoupled}, SLCNet  \cite{liu2022semi}, SCPNet  \cite{zhang2023self}, MCF  \cite{wang2023mcf}, BCP  \cite{bai2023bidirectional}, ABD \cite{chi2024adaptive}, GA-BCP \cite{wu98gradient}, AD-MT \cite{zhao2025alternate}.  As shown in Tab. \ref{table:compare with sota}, we achieve significant improvements on 2D datasets. What is particularly encouraging is that in both of these 2D datasets, we achieve superior performance while using only 5\% labeled samples, surpassing methods that use 10\% labeled samples. This further underscores the outstanding performance of our approach. As shown in Fig. \ref{fig:2D}, we can clearly observe that when dealing with smaller target organs. This is capable of accurately predicting these target organs in their entirety while other methods may fail to predict successfully or produce misleading results.  

\subsection{Ablation Studies}
\begin{table*}[!t]

	\resizebox{\linewidth}{!}{
		\begin{tabular}{ccccccccc}
			\toprule
			&{No.}  &{Pre-warm}  &{P$^3$M}   &{$\mathcal{L}_{bf}$}  &Dice    &Jaccard    &HD95    &ASD 	  \\
			\midrule
   \multirow{3}*{ACDC}		     &1                   &\checkmark       &                                 &                                 &60.58  &49.64         &26.97    &11.59                  \\		
			     &2                                                        &\checkmark		 &\checkmark	      &                                 &87.58  & 78.56        &2.85      & 1.13                 \\		
			     &3                                                        &\checkmark       &\checkmark         &\checkmark        &\textbf{89.71}    &\textbf{81.88}          &\textbf{1.85}    &\textbf{0.64}            \\
     \midrule
     \multirow{3}*{LA}		     &1                    &\checkmark       &                                 &                                  &83.87  &72.93          &15.72    &4.78                  \\		
			     &2                                                        &\checkmark		 &\checkmark	      &                                &88.59  &80.14        &7.95      & 2.59                 \\			
			     &3                   &\checkmark       &\checkmark         &\checkmark        &\textbf{89.57}    &\textbf{81.17}          &\textbf{7.83}    &\textbf{2.33}            \\
			
			\bottomrule
	\end{tabular}}
    \caption{Ablation study of effects of different components on ACDC and LA dataset with 5\% labeled data. Best results are in bold.}
	\centering
	\label{table:effects of different components}

\end{table*}
We conduct ablation studies to show the impact of each component in P$^3$Net  on the ACDC (2D) dataset and LA (3D) dataset. Including effects of different components, The curve of the  P$^3$M and the period of the  P$^3$M.

\noindent\textbf{Effects of Different Components.} 
We progressively add our components to validate the effectiveness of each one. On the ACDC and LA datasets with 5\% labeled data. As shown in the first row of 
 Tab. \ref{table:effects of different components}, we initially apply a basic  pre-warm stage to learn information from labeled samples without P$^3$M and the loss function. 
 Building on this pre-warm strategy, we then introduce P$^3$M to explore its fuction (second row). These demonstrate significant improvements upon incorporating P$^3$M, such as 20\% increase in Dice and 28.62\% increase in Jaccard on ACDC. This indicates that P$^3$M acts as a bridge between labeled and unlabeled data, transforming perturbations into positive guidance. 
 It inspires the network to learn more from unlabeled data, enhancing resistance to interference and overall robustness. To further explore the effectiveness of the loss function, we integrate it with P$^3$M. This results in 2.13\% increase in Dice and 3.32\% increase in Jaccard on ACDC, confirming that our loss function more effectively guides the network through a slow and periodic interpolation process, indirectly improving the network's performance.

\begin{table*}[!t] 

 \resizebox{\linewidth}{!}{
		\begin{tabular}{c|lllll}
\hline
\multicolumn{1}{l|}{} & Curve                                            & Dice  & Jaccard & HD95  & ASD  \\ \hline
\multirow{7}{*}{ACDC} & \textcolor{green}{green}                   & 84.11 & 74.22   & 2.88  & 0.8  \\
                      & \textcolor{purple}{purple}             & 86.00 & 76.86   & 9.82  & 3.81 \\
                      & \textcolor{blue}{blue}                        & 88.88 & 80.56   & 2.9   & 0.74 \\
                      & \textcolor{red}{red}              & \textbf{89.71} & \textbf{81.88}   & \textbf{1.85}  & \textbf{0.64} \\
                      & black & 86.82 & 77.90    & 9.33  & 3.14 \\
                      & \textcolor{yellow}{yellow}                   & 84.99 & 74.73   & 4.21  & 1.23 \\
                      & \textcolor{cyan}{cyan}                                               & 85.82 & 76.18   & 6.12  & 1.89 \\ \hline
\multirow{7}{*}{LA}   & \textcolor{green}{green}                   & 85.45 & 75.63   & 17.65 & 5.27 \\
                      & \textcolor{purple}{purple}             & 86.54 & 76.58   & 15.56 & 6.85 \\
                      & \textcolor{blue}{blue}                        & 88.85 & 80.02   & 9.9   & 3.11 \\
                      & \textcolor{red}{red}              & \textbf{89.57} & \textbf{81.17}   & \textbf{7.83}  & \textbf{2.33}\\
                      & black & 87.53 & 77.65   & 12.59 & 4.65 \\
                      & \textcolor{yellow}{yellow}                   & 85.03 & 75.32   & 17.09 & 6.32 \\
                      & \textcolor{cyan}{cyan}                                              & 85.12 & 75.98   & 18.13 & 5.79 \\ \hline
\end{tabular} 
	}
	\caption{Ablation study of the curve of the P$^3$M on ACDC(5\%) and LA(5\%) datasets. In the Fig. \ref{fig:down}, the specific function curve is as shown.
}
	
	\centering
    \label{table:temporal curve}
\end{table*}

\begin{figure*}
    \centering

    \centering
    \scalebox{0.9}{
    \includegraphics[width=\linewidth]{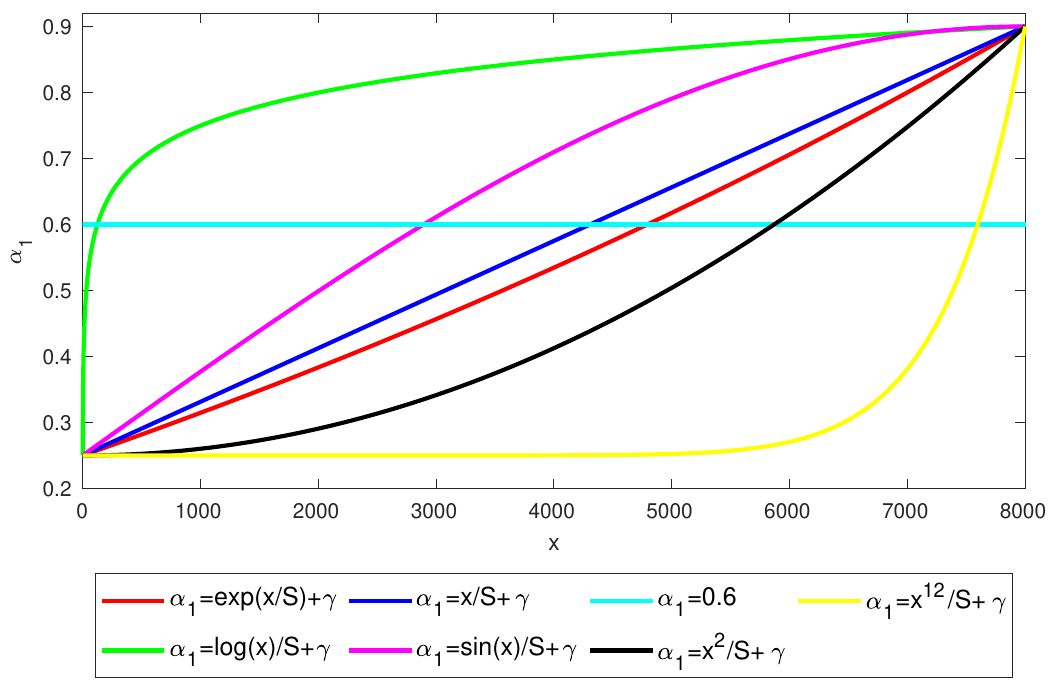}} 
    
    \caption{Demonstrate a period of different P$^3$M curve functions.  (Corresponding to each row in Tab.  \ref{table:temporal curve}.)}
    \label{fig:down}
  \label{fig14}
\end{figure*}
\begin{table}[]
    
	\resizebox{\linewidth}{!}{
		\begin{tabular}{l|lcccc}
		\toprule
		&{Period}              &Dice    &Jaccard  &HD95 &ASD       \\		
		\midrule
		\multirow{5}*{ACDC}&$T=4000$   &89.07&80.94&3.09&1.10\\
		&$T=8000$(our)  &\textbf{89.71}    &\textbf{81.88}   &1.85    &0.64\\
		&$T=12000$          &89.54&81.60&\textbf{1.63}&\textbf{0.55}\\ 
            &$T=16000$          &89.24&81.16&2.36&0.67\\
            &$T=75000$          &88.52&72.32&2.27&0.72\\
            \midrule
            \multirow{5}*{LA}&$T=4000$  &89.14&80.49&6.86&2.09 \\
		&$T=8000$(our)  &\textbf{89.57}    &\textbf{81.17}          &\textbf{7.83}    &\textbf{2.33}\\
		&$T=12000$          &88.99&80.25&9.70&3.17\\
            &$T=16000$          &87.17&77.34&41.42&11.68\\
            &$T=75000$          &86.22&76.03&15.99&3.95\\
  
		\bottomrule
	\end{tabular}}
    \caption{ Ablation study of the period of  the P$^3$M on the ACDC(5\%) and LA(5\%) dataset. $T=75000$ represents only one  complete training period.}
	\centering
        
	\label{table: period of the interpolation function}
\end{table}
\noindent\textbf {Curve  of  P$^3$M.} 
To validate the effectiveness of our progressive perturbation approach, we conduct various perturbative function and assess their impact on performance. As showed in the Tab. \ref{table:temporal curve} and Fig. \ref{fig:down}, the green curve represents a scenario where the interpolation ratio is initially small, but rapidly increases to a higher interpolation level over a short period. 
In this case, the network encounters fewer unlabeled data and predominantly learns previously acquired information, leading to rapid overfitting.
In contrast, the yellow curve depicts a scenario where the interpolation ratio remains low for most of the time but sharply rises towards the end of the cycle. These curves spend a majority of their time at a low interpolation level, introducing more disturbances to the network and impeding its ability to effectively utilize labeled data for guiding the learning of unlabeled data, thus hindering normal network learning. Both of these scenarios result in subpar performance.
The remaining four interpolation functions (purple, blue, red, black) all exhibit good performance. As seen from Table \ref{table:temporal curve}, they notably outperform the two extreme cases. Of particular note is the red curve we employed, inspired by the human learning memory curve, which demonstrated superior results in our model's progressive learning process. Additionally, we experimented with several other typical functions, all of which performed well, affirming the effectiveness of our adopted progressive periodic learning approach. It's worth emphasizing that while there may be "memory curves" better suited for the model in the future, our progressive learning mechanism remains crucial.

\begin{table}[]
     
	\resizebox{\linewidth}{!}{
		\begin{tabular}{ll|lllll}
\toprule
\multicolumn{2}{c|}{\multirow{2}{*}{Dice}}                & \multicolumn{5}{c}{Lower bounds}                                          \\ \cline{3-7} 
\multicolumn{2}{c|}{}                                     & 0.15  & 0.2   & 0.25           & 0.3   & 0.35                      \\ \hline
\multicolumn{1}{l|}{\multirow{5}{*}{Upper bounds}} & 0.8  & 87.79 & 88.62 & 89.69          & 88.96 & 89.62                     \\
\multicolumn{1}{l|}{}                              & 0.85 & 87.52 & 88.64 & 89.57          & 89.66 & 89.64                     \\
\multicolumn{1}{l|}{}                              & 0.9  & 88.80 & 89.55 & \textbf{89.71} & 89.65 & 89.57                     \\
\multicolumn{1}{l|}{}                              & 0.95 & 86.21 & 85.88 & 87.54          & 88.74 & 88.63                     \\
\multicolumn{1}{l|}{}                              & 1.0  & 85.69 & 86.12 & 88.51          & 86.82 & \multicolumn{1}{c}{86.21} \\ \bottomrule

\end{tabular}
 }
 \caption{ Ablation study of upper and lower bounds of P$^3$M on the ACDC(5\%)  dataset. }

	\centering
	\label{table: upper and lower bounds}
\end{table}
\noindent\textbf{Period of P$^3$M.} 
We use periodic learning to gradually and systematically acquire new information over time, refine understanding, and even temporary setbacks that may occur intermittently. 
Since we employ a method with fixed upper and lower bounds, the length of the period not only determines the number of interpolations within each cycle but also affects the slope of the interpolation function. A shorter periodic results in a steeper slope and faster changes in the interpolation function, while a longer period leads to a shallower slope and slower changes. Therefore, selecting an appropriate period is crucial. As shown in Tab. \ref{table: period of the interpolation function}. Initially, we conduct an experiment with a complete cycle of T=75000, treating the entire training process as one cycle. Subsequently, we carry out a series of experiments with T values of 4000, 8000, 12000, and 16000. We find that T=8000 yielded the best performance in both 2D and 3D datasets. While different periods do have some impact on the curves, our performance has been improved compared to the single period (T=75000). This also demonstrates the necessity of our periodic approach.

\noindent\textbf{Upper and lower bounds of P$^3$M}.
The variation of upper and lower bounds can adjust the strength of perturbations, thereby controlling the quality of those perturbations. With a fixed period, changes in the bounds will affect the values of 
$S$ and $\gamma$. Excessive interpolation may lead to unnecessary noise, exceeding the model's generalization capabilities, while insufficient interpolation might hinder the extraction of desired learning information. Therefore, selecting appropriate upper and lower bounds is crucial. In Table \ref{table: upper and lower bounds}, the first row represents the lower bound, and the first column represents the upper bound. When the upper bound is set to 0.9 and the lower bound is set to 0.25, performance is optimal. Slightly decreasing the upper bound or slightly increasing the lower bound may result in a minor performance decrease, but the impact is minimal. However, setting the lower bound to 0 will reduce the amount of unlabeled data the network learns, leading to overfitting to labeled data. Conversely, excessively increasing the upper bound will increase the proportion of unlabeled data, introduce more noise, and result in decreased performance.
\begin{figure}[t]
    \centering

    \centering
   
     \includegraphics[width=1\linewidth]{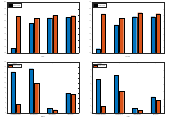}
    \caption{Application of P$^3$M.  Results are evaluated on the ACDC(5\%) and LA(5\%) dataset. Blue represents without our P$^3$M, and orange represents with our P$^3$M. (best viewed by zoom-in on screen.)}
    \label{fig:table5}
\end{figure}
\subsection{Applications}

\textbf{Application of P$^3$M.} We apply the proposed P$^3$M in two top-ranking models (BCP  \cite{bai2023bidirectional} and ClassMix  \cite{olsson2021classmix}) to achieve improved effectiveness. BCP and Classmix employ different sampling approaches: BCP utilizes bidirectional copy-paste for interpolation, while Classmix randomly selects half of the classes from one image and interpolates them onto another. 
    
    Both methods use fixed ratio interpolation. We apply our P$^3$M to them while keeping their mask M shapes unchanged. We adjust their interpolation ratios according to our mechanism. In Fig. \ref{fig:table5}, we show the quantitative comparison of the original models and the improved models(+ P$^3$M), it can be observed that our  P$^3$M largely improves the efficiency of original models. Especially at the Classmix method,  we achieved a remarkable 28.30\% improvement in Dice scores on the ACDC dataset when combined with our P$^3$M.
\begin{figure}
    \centering
    \includegraphics[width=1\linewidth]{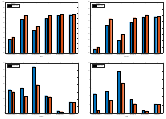}
    \caption{Application of  boundary-focused loss. Results are evaluated on the ACDC(5\%) and LA(5\%) dataset. Blue represents without our boundary-focused loss, and orange represents with our boundary-focused loss. (best viewed by zoom-in on screen.)}
    \label{fig:table6}
\end{figure}

\textbf{Application of Boundary-Focused Loss.} We apply our boundary-focused loss to various interpolation methods to assess its effectiveness in helping the network correct errors. As shown in the Fig. \ref{fig:table6}, we selected three of the latest methods to validate our loss function. We apply our loss function to other methods in order to compare the changes in their performance. We found that after introducing our loss function, their performance improved. This also indicates that our loss function, to a certain extent, reduces prediction errors caused by the loss of contextual information, indirectly enhancing the overall network performance.

\section{Conclusion}
In this paper, we strive to embrace challenges towards Interpolation-based perturbation for SSLMIS. We propose a new progressive and periodic perturbation mechanism which systematically regulates interpolated perturbations, acting as a bridge between labeled and unlabeled data to transform positive guidance in a more coherent manner aligned with human learning principles,thereby, enhancing the models' learning ability. We further propose a boundary-focused loss to improve the networks’s confidence for interpolation edge prediction errors caused by the lack of contextual information at the edges. Experimental results demonstrate that our proposed method can achieve state-of-the-art performance on two 2D and two 3D dataasets.
Additionally, we applied P$^3$M and the boundary-focused loss to SSLMIS perturbation methods, and our experiments demonstrated significant improvements in model performance.
\bibliographystyle{unsrt}
\bibliography{main}

\end{document}